\documentclass[12pt]{article}
\usepackage{amsmath}
\usepackage{amsfonts}
\usepackage{amssymb}

\usepackage{citesort}

\newtheorem{theorem}{Theorem}[section]

\newtheorem{Proposition}[theorem]{Proposition}

\newtheorem{remark}[theorem]{Remark}

\newcommand{\loc}{{
\rm \scriptsize loc}}

\numberwithin{equation}{section}

\DeclareFontFamily{OT1}{rsfs}{} \DeclareFontShape{OT1}{rsfs}{m}{n}{
<-7> rsfs5 <7-10> rsfs7 <10-> rsfs10}{}
\DeclareMathAlphabet{\mycal}{OT1}{rsfs}{m}{n}

\newcommand{\beaa}{\begin{eqnarray*}}
\newcommand{\eeaa}{\end{eqnarray*}}

\newcommand{\bel}[1]{\begin{equation}\label{#1}}
\newcommand{\bea}{\begin{eqnarray}}
\newcommand{\bean}{\begin{eqnarray}\nonumber}
\newcommand{\beal}[1]{\begin{eqnarray}\label{#1}}
\newcommand{\eea}{\end{eqnarray}}

\newcommand{\be}{\begin{equation}}
\newcommand{\eeq}{\end{equation}}
\newcommand{\ee}{\end{equation}}
\newcommand{\beqa}{\begin{eqnarray}}
\newcommand{\eeqa}{\end{eqnarray}}
\newcommand{\beqan}{\begin{eqnarray*}}
\newcommand{\eeqan}{\end{eqnarray*}}
\newcommand{\ba}{\begin{array}}
\newcommand{\ea}{\end{array}}

\newcommand{\mcM}{{\mycal M}}

\newcommand{\ii}{|I|}

\newcommand{\Scri}{{\mycal I}}

\newcommand{\R}{\mathbb R}

\newcommand{\mF}{{\cal F}}

\newcommand{\gini}{\bar g}
\newcommand{\Rini}{\bar R}
\newcommand{\Kini}{K}
\newcommand{\kini}{\Kini}
\newcommand{\Aini}{\bar A}
\newcommand{\Eini}{\bar E}

\newtheorem{Theorem} {\sc  Theorem\rm} [section]

\newtheorem{theo} [Theorem] {\sc  Theorem\rm}

\newcommand{\val}[1]{|#1|}

\newcommand{\hyp}{\mycal S}

\newcounter{mnotecount}[section]

\newcommand{\mnote}[1]
{\protect{\stepcounter{mnotecount}}$^{\mbox{\footnotesize
$
\bullet$\protect\themnotecount}}$ \marginpar{
\raggedright\tiny\em $\!\!\!\!\!\!\,\bullet$\protect\themnotecount:
#1} }

\renewcommand{\themnotecount}{\thesection.\arabic{mnotecount}}

\newcommand{\eq}[1]{(\ref{#1})}

\begin{document}

\title{Global solutions of the Einstein-Maxwell equations
in higher dimensions}
\author{Yvonne Choquet-Bruhat\\Acad\'emie des Sciences, Paris, and \\Albert Einstein Institute, Golm
\and Piotr T. Chru\'{s}ciel, Julien Loizelet\\F\'{e}d\'{e}ration Denis Poisson, LMPT, Tours, and \\Albert Einstein Institute, Golm}
\maketitle
\begin{abstract}
  We consider the Einstein-Maxwell equations in space-dimension $n$. We point out that the Lindblad-Rodnianski stability proof
  applies to those equations whatever the space-dimension $n\ge 3$. In even space-time dimension $n+1\ge 6$
  we use the
  standard conformal method on a Minkowski background to give a simple proof
  that the maximal globally hyperbolic development of initial data sets
  which are sufficiently close to the data for Minkowski space-time
  and which are Schwarzschildian outside of a compact set lead to
  geodesically complete space-times, with a complete Scri, with smooth conformal
  structure, and with the gravitational field approaching the Minkowski metric along null
  directions
 at least as fast as $r^{-(n-1)/2}$.
\end{abstract}

\section{Introduction}

There is increasing interest in asymptotically flat solutions of
Einstein equations in higher dimensions, see
e.g.~\cite{HollandsWald,HollandsIshibashi,BCS,EmparanReallReview}.
The pioneering work of Christodoulou and Klainerman
\cite{ChristodoulouKlainerman93} proving the nonlinear stability of
four-dimensional Minkowski spacetime uses the Bianchi equations, and
therefore does not extend to dimensions larger than four in any
obvious way. Now, global existence on $\R^{n+1}$ with $n\geq4$ for
small initial data of solutions of quasi-linear wave equations of
the type of Einstein's equations in wave coordinates has been proved
in~\cite{LiChen,HormanderGlobal}%
\footnote{Those works build upon
\cite{KlainermanGlobalCPAMTwo,Klainerman:null}; however the
structure conditions
in~\cite{KlainermanGlobalCPAMTwo,Klainerman:null} are not compatible
with
the Einstein equations.}%
, see also~\cite{Christodoulou:global} for
odd $n\ge 5$, but the analysis there assumes fall-off of initial
data near spatial infinity incompatible with the Einstein
constraints%
\footnote{In~\cite{LiChen,HormanderGlobal} compactly supported
data are considered. In the theorem for general quasilinear
systems given in~\cite{Christodoulou:global} the initial data are
in a Sobolev space which requires fall-off at infinity faster than
$r^{-n-3/2}$. This should be compared with a fall-off  of
$g_{\mu\nu}-\eta_{\mu\nu}$ \emph{not faster} than $r^{-n+2}$
required by the positive energy theorem.}%
.

In this note we point out that the Lindblad-Rodnianski stability
argument~\cite{LindbladRodnianski,LindbladRodnianski2} in
space-dimension $n=3$ can be repeated for all $n\ge 3$ for the
Einstein-Maxwell system. Thus, Minkowski space-time is indeed stable
against electro-vacuum non-linear perturbations in all dimensions
$n+1\ge 4$.

Next, we point out that non-linear electro-vacuum stability, for
initial data which are Schwarzschildian outside of a compact set,
can be proved by the standard conformal method on Minkowski
space-time for odd $n\ge 5$. As usual, the method gives  detailed
information on the asymptotic behavior of the gravitational field,
not directly available in the Lindblad-Rodnianski method.

It should be mentioned that in \emph{vacuum}, and in even space-time
dimensions $n+1\ge 4$, existence of smooth conformal completions has
been proved in~\cite{AndersonChrusciel} using the Fefferman-Graham
obstruction tensor, for initial data which are stationary outside of
a compact set.  The argument there is simpler than the
Lindblad-Rodnianski method, but less elementary than the standard
conformal method presented here. Moreover, the direct conformal
method here provides more  information about the asymptotics of the
fields; however, for hyperboloidal initial data our conditions are
more restrictive. In any case, it is not clear whether the argument
of~\cite{AndersonChrusciel} generalizes to Einstein-Maxwell
equations. (Compare \cite{Friedrich,friedrich:JDG} for a completely
different approach when $n=3$.)

\section{Nonlinear stability in higher dimensions}

Consider the Einstein-Maxwell equations, in space-time dimension
$n+1$,
\begin{equation}\label{em1}\left \{ \begin{array}{ll} R_{\mu\nu}-
\frac{1}{2}g_{\mu\nu}R=8\pi T_{\mu\nu}\;;\\
D_{\mu}\mathcal{F}^{\mu\nu}=0\;,
\end{array} \right.\end{equation}
with
$T_{\mu\nu}=\frac{1}{4\pi}(\mathcal{F}_{\mu\lambda}\mathcal{F}_{\nu}^
{~\lambda}-\frac{1}{4}
g_{\mu\nu}\mathcal{F}^{\lambda\rho}\mathcal{F}_{\lambda\rho})$ and
$\mathcal{F}_{\mu \nu
}=\partial_{\mu}A_{\nu}-\partial_{\nu}A_{\mu}$. It is assumed
throughout that $n\ge 3$.

We suppose that we are given an $n$--dimensional Riemannian manifold
$(\hyp,\gini  )$, together with a symmetric tensor $\kini_{  }$, and
initial data  $(\Aini =\Aini _idx^i,\Eini =\Eini _idx^i)$ for the
Maxwell field. (Throughout this work, quantities decorated with a
bar
 are  pull-backs to the initial data surface $
\hyp$.)
For the stability results we will assume that $\hyp=\R^n$, but
this will not be needed for the local existence results. We seek a
Lorentzian manifold $(\mcM =\R\times \hyp,g)$ with a one-form
field $A$, satisfying \eq{em1}, such that $\gini  $ is the
pull-back of $g$, $\Kini $ is the extrinsic curvature tensor of
$\{0\}\times\hyp$, while $(\Aini ,\Eini )$ are the pull-backs to
$\hyp$ of the vector potential $A_\mu dx^\mu$ and of the electric
field $\mF_{\mu\nu} dx^\mu n^\nu$, where $n^\mu$ is the field of
unit normals to $\hyp$.

We assume the following fall-off behaviors as  $r=\val{x} $ tends to
infinity, for some constants $\alpha > 0$,  $m$:
\begin{equation}\label{ci}\forall~ i,j=1,...,n~~~~
\left \{ \begin{array}{ll}
 \gini_{ij}=\begin{cases}(1+\frac{2m}{r})\delta_{ij}+O(r^{-1-\alpha})\;,\;
  & \text{ for }n=3\;,\\
 \delta_{ij}+O(r^{-\frac{n-1}{2}-\alpha})\;,\;
  &\text{ for }n\geq 4\;,\end{cases}\\
 \Aini _i=O(r^{-\frac{n-1}{2}-\alpha})\;,
 \\
   \kini_{  ij}=O(r^{-\frac{n+1}{2}-\alpha})\;,\\
    \Eini _i=O(r^{-\frac{n+1}{2}-\alpha})\;. \\
\end{array} \right.
\end{equation}
We will, of course, assume that the constraint equations hold:
\begin{equation}\label{con}
\forall~ i,j=1,...,n ~~~~\left \{ \begin{array}{ll} \Rini-\kini_{
j}^{i}\kini_{  i}^{j}+\kini_{  i}^{i}\kini_{  j}^{j}= 2\mF
_{0i}\mF ^{~i}_{0}+\mF _{ij}\mF ^{ij}\;,
\\     D^{j}\kini_{  ij}-D_{i}\kini_{  j}^{j}=\mF _{0j}\mF _i^{~j}
~~\;,
\\\nabla_i\mF ^{0i}=0\;.
 \end{array} \right.\end{equation}
where $\Rini$ is the scalar curvature of  $\gini  $, $D$ is the
covariant derivative operator associated with $\gini  $ while
$\nabla$ is the space-time covariant derivative.

The following result can be proved by a repetition of the arguments
in~\cite{LindbladRodnianski,LindbladRodnianski2}. We refer the
reader to~\cite{LoizeletCRAS,Loizelet:these} for details and for
some information on the asymptotic behavior of the fields:

\begin{theo}\label{main}
Let $(\hyp=\R^{n },\gini ,\kini ,\Aini ,\Eini )$, $n\ge 3$, be
initial data for the Einstein-Maxwell equations \eq{em1} satisfying
\eq{ci} and \eq{con}, with ADM mass $m$, set
  \bel{Nn}
N_n
= 6+2\left[\frac{n+2}{2}\right]
 \;.
 \ee
Write $\gini =\delta+h_0^0+h_0^1\;$ with
$$
 h^0_{0ij}=\begin{cases}\chi
 (r)\frac{2M}{r}\delta_{ij}\quad &\text{for }n=3\;,
  \\0\quad&\text{for
 }n\geq 4\;,\end{cases}
 $$
 for a function $\chi\in C^\infty$ equal to $ 1$ for $r\geq 3/4$ and to $0$ for $r\leq 1/2$.
Define\footnote{We take this opportunity to note a misprint in the
 norm in~\cite[Eq.~(9)]{LoizeletCRAS}.}
\begin{align}
E_{N_n,\gamma}(0)=&\sum_{0\leq\ii\leq N_n}\left(
\val{\val{(1+r)^{1/2+\gamma+\ii}\nabla\nabla^Ih_0^1}}_{L^2}^2 +
\val{\val{(1+r)^{1/2+\gamma+\ii}\nabla^I\kini }}_{L^2}^2
\right.\\\nonumber
&\left.+\val{\val{(1+r)^{1/2+\gamma+\ii}\nabla\nabla^I \Aini
}}_{L^2}^2 +\val{\val{(1+r)^{1/2+\gamma+\ii}\nabla^I \Eini
}}_{L^2}^2 \right)\;.
\end{align}
There exist constants $\varepsilon_0>0$ and
$\gamma_0(\varepsilon_0)$, with $\gamma_0(\varepsilon_0)\rightarrow
0 $ as $\varepsilon_0\rightarrow 0$, such that, for all initial data
satisfying
\begin{equation}\label{small}
\sqrt{E_{N_n,\gamma}(0)}+m\leq\varepsilon_0\;, \end{equation}
 for a certain $\gamma>\gamma_0$, the Cauchy problem described above has a global solution
 $(g,A)$ defined on $\R^{n+1}$,
with $(\mathbb{R}^{n+1},g)$ --- geodesically complete.
The solution is smooth if the initial data are.
\end{theo}

The threshold value  $N_n$ in \eq{Nn} arises, essentially, from the
$n$-dimensional Klainerman-Sobolev inequalities~\cite{Hormander97}.

 For initial data which are polyhomogeneous, or
conformally smooth, at $i^0$, one expects that the solutions will
have a polyhomogeneous conformal completion at $\Scri$. 
Unfortunately, the information about the asymptotic
behavior of the fields obtained in the course of the proof of
Theorem~\ref{main} does not establish this. In the remainder of
our work we address this question, in odd space-dimension $n\ge
5$, for a restricted class of initial data. We will use a
conformal transformation to both give an alternative, simpler
proof of global existence for small data, and obtain information
on the asymptotic behavior.

\section{Cauchy problem for the vacuum Einstein equations in wave coordinates}

We first recall some well known facts. We start with the  Einstein
equations without sources, the Einstein-Maxwell equations are
considered in Section~\ref{SEM} below.

The vacuum Einstein equations in wave coordinates on $\R^{n+1}$
constitute a set of quasi-diagonal quasi-linear wave equations for
the components $g_{\mu\nu}$ of the spacetime metric $g$, which we
write symbolically as
\begin{equation}
g^{\alpha\beta}\frac{\partial^{2}g }{\partial x^{\alpha}\partial
x^{\beta}}=F (g)(\frac{\partial g}{\partial x})^{2}\;,
\end{equation}
where  the right-hand-side is a  quadratic form in the first
derivatives of the $g $'s with coefficients polynomials in the $g
$'s and their contravariant associates. We take $\R^{n}$ as an
initial manifold. The spacetime manifold will be a subset $V$
  of
$\R^{n+1}$. The geometric initial data on $\R^{n}$ are a metric
$\bar{g}$ and a symmetric 2-tensor $\kini $.

We suppose also given on $\R^{n}$ the lapse $\bar{N}$, shift
$\bar{\beta}$ as well as their time derivatives, chosen so that the
corresponding spacetime metric and its first derivatives satisfy on
$\R^{n}$ the harmonicity conditions $\overline{F^{\mu}}=0$. If the
initial data $\bar{g},K$ satisfy the constraints, a solution of the
Einstein equations in wave coordinates on $V$ is such that the
harmonicity functions $F^{\mu}:=\Box_g x^\mu$ satisfy an homogeneous
linear
system of wave equations on $(V,g)$, with also $\overline{\partial_{t}F^{\mu}%
}=0$, hence $F^{\mu}=0$ if $(V,g)$ is globally hyperbolic.

\section{Conformal mapping}

\subsection{Definition}

To prove our global existence result we use a mapping $\phi:x\mapsto
y$ from the future timelike cone with vertex $0$,
$I_{\eta,x}^{+}(0)$, of a Minkowski spacetime, which we denote
$(\R_{x}^{n+1},\eta_{x})$, into the past timelike cone with vertex
$0$
of another Minkowski spacetime, $(\R_{y}^{n+1}%
,\eta_{y})$. This map is defined by, with $\eta$ the diagonal
quadratic form ($-1,1,...,1)$,
\begin{equation}
\phi:\text{ \ \ }I_{\eta,x}^{+}(0)\rightarrow \R_{y}^{n+1}\text{ \ \
by
\ }x^{\alpha}\mapsto y^{\alpha}:=\frac{x^{\alpha}}{\eta_{\lambda\mu}%
x^{\lambda}x^{\mu}}\text{ }.
\end{equation}
It is easy to check that $\phi$ is a bijection from $I_{\eta,x}^{+}(0)$ onto
$I_{y,\eta}^{-}(0)$ with inverse
\begin{equation}
\phi^{-1}:y\mapsto x\text{ \ by \ \ \ }x^{\alpha}:=\frac{y^{\alpha}}%
{\eta_{\lambda\mu}y^{\lambda}y^{\mu}}\;. \end{equation} Moreover
$\phi$ is a conformal mapping between Minkowski metrics, it holds
that
\begin{equation}
\eta_{\alpha\beta}dx^{\alpha}dx^{\beta}=\Omega^{-2}\eta_{\alpha\beta
}dy^{\alpha}dy^{\beta}%
\end{equation}
where $\Omega$ is a function defined on all $\R_{y}^{n+1}$, given by
\begin{equation}
\Omega:=\eta_{\alpha\beta}y^{\alpha}y^{\beta}\;.
 \end{equation}
This conformal mapping appears to be better adapted to the context
of the Einstein equations, which involve constraints, than the
Penrose transform used in~\cite{Christodoulou:global} for general
quasi-linear wave equations.

\subsection{Transformed equations}

We set
\begin{equation}
f_{\mu\nu}:=g_{\mu\nu}-\eta_{\mu\nu}\;. \end{equation} We consider
$f:=(f_{\mu\nu})$ as a \emph{set of scalar functions} on
$\R_{x}^{n+1}. $ The Einstein equations in wave coordinates, for the
unknowns $f:=(f_{\mu\nu })$, are then a quasi-diagonal set of
quasi-linear wave equations of the form
\begin{equation}
\label{3.6}
\eta^{\alpha\beta}\frac{\partial^{2}f}{\partial x^{\alpha}\partial x^{\beta}%
}=-(g^{\alpha\beta}-\eta^{\alpha\beta})\frac{\partial^{2}f}{\partial
x^{\alpha}\partial x^{\beta}}+F(\eta+f)(\frac{\partial f}{\partial
x})^{2}\;. \end{equation}
The general relation between the wave operator on scalar functions
in two conformal metrics transforms the left-hand-side of \eq{3.6}
into the following partial differential operator (compare
Remark~\ref{R1} below)
\begin{equation}
\label{3.7}
\eta^{\alpha\beta}\frac{\partial^{2}(\Omega^{-\frac{n-1}{2}}f\circ\phi^{-1}%
)}{\partial y^{\alpha}\partial y^{\beta}}\equiv\Omega^{-\frac{n+3}{2}}%
(\eta^{\alpha\beta}\frac{\partial^{2}f}{\partial x^{\alpha}\partial x^{\beta}%
})\circ\phi^{-1}\;. \end{equation}
We introduce the following new set of scalar functions on $\R_{y}^{n+1}$%
\begin{equation}
\hat{f}:=\Omega^{-\frac{n-1}{2}}f\circ\phi^{-1}\;,\text{ \ \ i.e. \
\ \ }\hat
{f}_{\mu\nu}:=\Omega^{-\frac{n-1}{2}}f_{\mu\nu}\circ\phi^{-1}\;.
\end{equation} With this notation, the system \eq{3.7} reads
\begin{equation}
\label{3.8}
 \eta^{\alpha\beta}\frac{\partial^{2}\hat{f}}{\partial
y^{\alpha}\partial
y^{\beta}}=-\Omega^{-\frac{n+3}{2}}\{(g^{\alpha\beta}-\eta^{\alpha\beta
})\frac{\partial^{2}f}{\partial x^{\alpha}\partial x^{\beta}}-F(\eta
+f)(\frac{\partial f}{\partial x})^{2}\}\circ\phi^{-1}\;.
\end{equation} We note that $g^{\alpha\beta}-\eta^{\alpha\beta}$ is
a rational function of $f\equiv(f_{\mu\nu})$ with numerator a linear
function of $f$ and with denominator bounded away from zero as long
as $g_{\alpha\beta}$ is non degenerate. Therefore we have
\begin{equation}
(g^{\alpha\beta}-\eta^{\alpha\beta})\circ\phi^{-1}=\Omega^{\frac{n-1}{2}}%
\hat{h}^{\alpha\beta}\;, \end{equation} with $\hat{h}^{\alpha\beta}$
a rational function of $\hat{f}$ and $\Omega^{\frac{n-1}{2}}$, with
denominator bounded away from zero as long as $g_{\alpha\beta}$ is
non degenerate. Now,
\begin{equation}
\eta^{\alpha\beta}\frac{\partial^{2}\hat{f}}{\partial
y^{\alpha}\partial y^{\beta}} \text{
}=-\Omega^{-2}\hat{h}^{\lambda\mu}\{\frac{\partial ^{2}f}{\partial
x^{\lambda}\partial x^{\mu}}\}\circ\phi^{-1}+\Omega
^{-\frac{n+3}{2}}\{F(\eta+f)(\frac{\partial f}{\partial
x})^{2}\}\circ \phi^{-1}\;. \end{equation} We use the definitions of
$\Omega$ and of the mapping $\phi$ to compute the right-hand-side of
\eq{3.8}. It holds that:
\begin{equation}
\frac{\partial\Omega}{\partial
y^{\alpha}}=2\eta_{\alpha\beta}y^{\beta }=:2y_{\alpha}\;,
\end{equation}
and
\begin{equation}
A_{\mu}^{\alpha}:=\frac{\partial y^{\alpha}}{\partial
x^{\mu}}\circ\phi
^{-1}\equiv\Omega\delta_{\mu}^{\alpha}-2y^{\alpha}y_{\mu}\;.
\end{equation} We see that $A_{\mu}^{\alpha}$ is bounded on any
bounded set of $\R_{y}^{n+1}. $

Elementary calculus gives for an arbitrary function $f$ on
$\R_{x}^{n+1}$ the identities
\begin{equation}
\frac{\partial f}{\partial x^{\mu}}\circ\phi^{-1}=A_{\mu}^{\alpha}%
\frac{\partial(f\circ\phi^{-1})}{\partial y^{\alpha}}\;,
\end{equation} and
\begin{equation}
\frac{\partial^{2}f}{\partial x^{\lambda}\partial
x^{\mu}}\circ\phi
^{-1}=A_{\lambda}^{\beta}\frac{\partial}{\partial
y^{\beta}}\{A_{\mu}^{\alpha
}\frac{\partial(f\circ\phi^{-1})}{\partial y^{\alpha}}\}\;.
\end{equation}
Hence, by straightforward calculation,
\begin{equation}
\frac{\partial^{2}f}{\partial x^{\lambda}\partial x^{\mu}}\circ\phi
^{-1}=B_{\lambda\mu}^{\alpha\beta}\frac{\partial^{2}(f\circ\phi^{-1}%
)}{\partial y^{\alpha}\partial y^{\beta}}+C_{\lambda\mu}^{\alpha}%
\frac{\partial(f\circ\phi^{-1})}{\partial y^{\alpha}}%
\end{equation}
where the coefficients $B$ and $C$ are bounded on any bounded subset
of $\R_{y}^{n+1}$. They are given by
\begin{equation}
B_{\lambda\mu}^{\alpha\beta}:=A_{\mu}{}^{(\alpha}A{}^{\beta)}_{\lambda}\;,
\end{equation} i.e.
\begin{equation}
B_{\lambda\mu}^{\alpha\beta}
 \equiv
 \Omega^{2}\delta{}^{(\beta}_{\lambda}%
\delta_{\mu}^{\alpha)}
 -2\Omega(y^{(\beta}y_{\lambda}\delta_{\mu}^{\alpha)
}+y^{(\alpha}y_{\mu}\delta_{\lambda}^{\beta)})+4y^{\alpha}y^{\beta}y_{\lambda
}y_{\mu}%
\end{equation}
and
\begin{equation}
C_{\lambda\mu}^{\alpha}\equiv-2\Omega(y_{\lambda}\delta_{\mu}^{\alpha}+y_{\mu
}\delta_{\lambda}^{\alpha}+y^{\alpha}\eta_{\lambda\mu})+8y^{\alpha}y_{\lambda
}y_{\mu})\;. \end{equation} If we now set
$f\circ\phi^{-1}=\Omega^{k}\hat{f}$, we find:
\begin{equation}
\frac{\partial(f\circ\phi^{-1})}{\partial y^{\alpha}}\equiv\frac
{\partial(\Omega^{k}\hat{f})}{\partial y^{\alpha}}=\Omega^{k}\frac
{\partial\hat{f}}{\partial y^{\alpha}}+2k\Omega^{k-1}y_{\alpha}\hat{f}%
\end{equation}
and
\begin{equation}
\frac{\partial^{2}(f\circ\phi^{-1})}{\partial y^{\alpha}\partial y^{\beta}%
}\equiv\Omega^{k}\frac{\partial^{2}\hat{f}}{\partial y^{\alpha}\partial
y^{\beta}}+2k\Omega^{k-1}(y_{\beta}\frac{\partial\hat{f}}{\partial y^{\alpha}%
}+y_{\alpha}\frac{\partial\hat{f}}{\partial y^{\beta}})+k\Omega^{k-2}%
D_{\alpha\beta}\hat{f}
 \;,
\end{equation}
with
\begin{equation}
D_{\alpha\beta}:=4(k-1)y_{\alpha}y_{\beta}+2\eta_{\alpha\beta}\Omega\;.
\end{equation}
 The second term on the right-hand-side of \eq{3.8} is
\begin{equation}
\Omega^{-\frac{n+3}{2}}\left\{F(\eta+f)(\frac{\partial f}{\partial
x})^{2} \right\}\circ\phi^{-1}
 \equiv
 \Omega^{2k-2-\frac{n+3}{2}}
  F(\eta+\Omega^{\frac
{n-1}{2}}\hat{f})\Big[A_{\mu}^{\alpha}
(\Omega\frac{\partial\hat{f}}{\partial
y^{\alpha}}+2ky_{\alpha}\hat{f} )\Big]^{2}
 \;.
\end{equation}
 Now, $A^\alpha_\mu y_\alpha=-\Omega y_\mu$, and recalling that $k=\frac{n-1}{2}$,    the
 right-hand-side of the last equation can be rewritten as
\begin{equation}
 \Omega^{ \frac{n-5}{2}}
  F(\eta+\Omega^{\frac
{n-1}{2}}\hat{f})\Big[A_{\mu}^{\alpha}
\frac{\partial\hat{f}}{\partial y^{\alpha}}-(n-1)y_{\mu}\hat{f}
\Big]^{2}
 \;.
\end{equation}
This shows that this term   extends smoothly, as long as the metric
$\eta+\Omega^{\frac{n-1}{2}}\hat{f}$ is non degenerate, to a smooth
system on $\R_{y}^{n+1}$ if $n\geq5$ and $n$ is odd.

The first term at the right-hand-side of \eq{3.8} is
\beaa \lefteqn{
-\Omega^{-2}\hat{h}^{\lambda\mu}\Big\{B_{\lambda\mu}^{\alpha\beta}\frac
{\partial^{2}(f\circ\phi^{-1})}{\partial y^{\alpha}\partial y^{\beta}%
}+C_{\lambda\mu}^{\alpha}\frac{\partial(f\circ\phi^{-1})}{\partial y^{\alpha}%
}\Big\}\equiv}&&
 \\
 &&
\Omega^{k-4}\hat{h}^{\lambda\mu}\Big\{\Omega^{2}B_{\lambda\mu}^{\alpha\beta}%
\frac{\partial^{2}\hat{f}}{\partial y^{\alpha}\partial y^{\beta}%
}+[4kB_{\lambda\mu}^{\alpha\beta}\Omega y_{\beta}+\Omega^{2}C_{\lambda\mu
}^{\alpha}]\frac{\partial\hat{f}}{\partial y^{\alpha}}+[B_{\lambda\mu}%
^{\alpha\beta}kD_{\alpha\beta}+2k\Omega C_{\lambda\mu}^{\alpha}y_{\alpha}%
]\hat{f}\Big\}
 \;.
\eeaa
A similar analysis shows that this again extends smoothly if the
metric $\eta+\Omega^{\frac{n-1}{2}}\hat{f}$ is non degenerate, if
$n\geq5$, and if $n$ is odd.

\begin{remark}
\label{R1}
 The particular case of the conformal covariance of the
wave equation that we have used, the identity \eq{3.7}, results by
a straightforward computation, when $k=\frac{n-1}{2}$, from the
obtained identities, together with
\begin{equation}
\eta^{\alpha\beta}\frac{\partial\Omega}{\partial y^{\beta}}\frac
{\partial\Omega}{\partial y^{\alpha}}\equiv4\Omega,\text{ \ \ and \ \ }%
\eta^{\alpha\beta}\frac{\partial^{2}\Omega}{\partial
y^{\alpha}\partial y^{\beta}}\equiv2(n+1)\;. \end{equation}
\end{remark}

We have proved

\begin{Proposition}
\label{Pp1}
 The bijection $\phi$ transforms the Einstein equations in
wave coordinates on $I_{\eta,x}^{+}(0)$ into a quasi-diagonal,
quasi-linear system on $I_{\eta ,y}^{-}(0)$ which extends smoothly
to $\R_{y}^{n+1}$, as long as the metric
$\eta+\Omega^{\frac{n-1}{2}}\hat{f}$ is non degenerate, $n$ is odd
and satisfies $n\geq5$.
\end{Proposition}

Proposition~\ref{Pp1} immediately shows propagation of the decay
rate $\Omega^{\frac{n-1}{2}}$ near $\Scri$ for hyperboloidal initial
data which, in wave coordinates, admit a smooth compactification as
described above with this decay rate. Further, hyperboloidal initial
data in this class which are sufficiently close to the Minkowskian
ones lead to solutions which are global to the future of a
hyperboloidal surface by the usual stability results. In the
remainder of this paper we will show that Proposition~\ref{Pp1} can
also be used to construct solutions which are global both to the
future and to the past.

\section{The local Cauchy problem in $\R_{x}^{n+1}$}

\subsection{Initial data}

We consider the Cauchy problem on $\R_{x}^{n+1}$ for the Einstein
equations   in wave coordinates with initial data $(\bar{g},K)$ on a
manifold $\R^{n}$ embedded as a submanifold
$$
 M_{x}:=\{x^{0}=2\lambda\}
$$
of $\R_{x}^{n+1}$. We suppose these initial data satisfy the
Einstein constraints, and we choose the initial lapse and shift so
that the harmonicity conditions
are everywhere initially zero. Any globally hyperbolic solution $(\mathcal{V}%
_{x}\subset \R_{x}^{n+1},g)$ of the Einstein equations in wave
coordinates taking these initial data is then a solution of the full
Einstein equations. Such a solution exists if the initial data
belong to $H_{s+1}^{\loc}\times H_{s}^{\loc}$, $s>\frac{n}{2}$.

For our global existence theorem we suppose that the initial data
coincide with a time-symmetric Schwarzschild initial data set
outside a ball $B_{R_x}$ of radius $r=R_x$. Large families of such
initial data sets have been constructed
in~\cite{ChDelay,CorvinoAHP},  arbitrarily close to those for
Minkowski space-time. (The construction of such data there is done
for time-symmetric initial data sets, but there is no \emph{a
priori} reason known why all such initial data sets should have
vanishing $\kini$.\footnote{A trivial example of initial data set
with non-zero $\kini$ is obtained by moving the initial data surface
in a time-symmetric space-time,  Schwarzschildian outside of a
compact set. This example raises the interesting question: is it
true that vacuum, maximal globally hyperbolic space-times which
contain a Cauchy surface which is Schwarzschildian at infinity
necessarily contain a totally geodesic Cauchy surface?})
 We will use this fact as in
previous works~\cite{ChDelay2,CutlerWald,AndersonChrusciel}, but
here with the Schwarzschild metric in wave coordinates.

 By the general
uniqueness theorem for quasi-linear wave equations, the solution
$(\mathcal{V}_{x},g)$ coincides with the Schwarzschild spacetime, in
wave coordinates, in the domain of dependence of the Schwarzschild
region, $M_{x}\backslash B_{R_x}$.

\subsection{The $(n+1)$--dimensional Schwarzschild metric in wave coordinates}
\label{sSwc}

The Schwarzschild metric $g_{m}$ with mass parameter $m$, in any dimension
$n\geq3$ is in standard coordinates
\begin{equation}
g_{m}=-\left(  1-\frac{2m}{\bar{r}^{n-2}}\right)  dt^{2}+\frac{d\bar{r}^{2}%
}{1-\frac{2m}{\bar{r}^{n-2}}}+\bar{r}^{2}d\Omega^{2}%
\end{equation}
where $d\Omega^{2}$ is the round unit metric on $S^{n-1}$.
Introduce a new coordinate system $x^{i}=r(\bar{r})n^{i}$, with
$n^{i}\in S^{n}$; the requirement that $x^{\mu}=(t,x^{i})$ be wave
coordinates, $\Box_{g}x^{\mu}=0$, is equivalent to the equation
\[
\frac{d}{d\bar{r}}[\bar{r}^{n-1}[1-\frac{2m}{\bar{r}^{n-2}}]\frac{dr}{d\bar
{r}}]={(n-1)\bar{r}^{n-3}}{r}\;.
\]
Setting $\rho=1/\bar{r}$, one obtains an equation with a Fuchsian singularity
at $\rho=0$:
\[
\frac{d}{d\rho}[\rho^{3-n}(1-{2m}{\rho^{n-2}})\frac{dr}{d\rho}]={(n-1)\rho^{1-n}%
}{r}\;.
\]
The characteristic exponents are $-1$ and $n-1$ so that, after matching a few
leading coefficients, the standard theory of such equations provides solutions
with the behavior
\[
r=\bar{r}-\frac{m}{(n-2)\bar{r}^{n-3}}+\left\{
\begin{array}
[c]{ll}%
\frac{m^{2}}{4}\bar{r}^{-3}\ln\bar{r}+O(\bar{r}^{-5}\ln\bar{r}), & n=4\\
O(\bar{r}^{5-2n}), & n\geq5
\end{array}
\right.
\]
Somewhat surprisingly, we find logarithms of $\bar{r}$ in an
asymptotic expansion of $r$ in dimension $n=4$. However, for
$n\geq5$ there is a complete expansion of $r-\bar{r}$ in terms of
inverse powers of $\bar{r}$, without any logarithmic terms: if we
write $g_{m}$ in the coordinates $x^{i}$, then
\begin{equation}
(g_{m})_{\mu\nu}=\eta_{\mu\nu}+(f_{m})_{\mu\nu}%
\end{equation}
\ with the functions $(f_{m})_{\mu\nu}$ of the form
\begin{equation}
(f_{m})_{\mu\nu}=\frac{1}{r^{n-2}}h_{\mu\nu}(\frac{1}{r},\frac{\vec{x}}%
{r}),\text{ \ \ }\vec{x}:=(x^{i})\;,
\end{equation}
with
$h_{\mu\nu}(s,\vec{w})$ analytic functions of their arguments near
$s=0$. In fact, there exist functions $h_{00}(s)$, $h(s)$, and
$\hat{h}(s)$, analytic near $s=0$, such that
\[
g_{m} =\Big(-1+\frac{h_{00}(r^{-1})}{r^{n-2}}\Big)(dx^{0})^{2}%
+\Big[(1+\frac{h(r^{-1})}{r^{n-2}})\delta_{ij}+\frac{\hat{h}(r^{-1})}{r^{n-2}%
}\frac{x^{i}x^{j}}{r^{2}}\Big]dx^{i}dx^{j}\;.
\]
As $\R^n$ is spin, we will necessarily have $m\ge 0$ for the initial
data that we consider.

\subsection{Domain of dependence of the Schwarzschild initial data}

The boundary $\mathcal{\dot{D}}_{x}%
^{+}(M_{x}\backslash B_{R_{x}})$ of the future domain of dependence $\mathcal{ {D}}_{x}%
^{+}(M_{x}\backslash B_{R_{x}})$ is threaded by null radial outgoing
geodesics of the Schwarzschild metric issued from $\dot{B}_{R_{x}}$,
solutions the differential equation
\begin{equation}
\frac{dt}{dr}=\Big(\frac{g_{m,rr}}{g_{m,tt}}\Big)^{\frac{1}{2}}%
\end{equation}
such that $t(R_{x})=2\lambda$, i.e.
\begin{equation}
\label{5.5}
x^{0}\equiv t=2\lambda+\int_{R_x}^{r}\Big(\frac{g_{m,rr}}{g_{m,tt}}\Big)^{\frac{1}{2}%
}dr\;. \end{equation} For the Schwarzschild metric we have
\begin{equation}
\Big|\Big(\frac{g_{m,rr}}{g_{m,tt}}\Big)^{\frac{1}{2}}-1\Big|
 =
 \Big| \frac{2m}{r^{n-2}(1-\frac{2m}{r^{n-2}})} \Big|
 \leq
\frac{4m}{r^{n-2}}
 \end{equation}
for $r\ge R_x$ provided that
\bel{mbound} \frac{2m}{R_x^{n-2}}\le 1/2
 \;;
 \ee
  at fixed $R_x$ this can be achieved
by requiring $m$ to be sufficiently small; alternatively at given
$m$ we can increase $R_x$.
 For
$n\geq5$  we deduce from \eq{5.5} that on
$\mathcal{\dot{D}}_{x}^{+}(M_{x}\backslash B_{R_{x}})$ we have
\begin{equation}
b\geq t-r\geq a\;. \end{equation} with
\begin{equation}
b:=2\lambda-R_x+\frac{4m}{(n-3)R_{x}^{n-3}},\text{ \ \ \
}a:=2\lambda -R_{x}-\frac{4m}{(n-3)R_{x}^{n-3}}\;. \end{equation}
We can always choose $\lambda$ large enough so that
$a>0$, requiring
 e.g. $2\lambda>R_x+1$ for $R_x\ge 1$ (assuming \eq{mbound}), which implies that $\dot {\mathcal{D}}^{+}(M_{x}\backslash
B_{R_{x}})$
is interior to $I_{\eta ,x}^{+}(0)$.

\section{The global Cauchy problem}

The local Cauchy problem for the system \eq{3.8} on
$M_{y}:=\{y^{0}=-\frac {1}{2\lambda}\}$, with initial data in
$H_{s+1}\times H_{s}$, $s>\frac{n}{2}$ in a ball of radius
$R_{y}>\frac{1}{2\lambda}$, and $\eta+\hat{f}_{|M_{y}}$ a non
degenerate Lorentzian metric, has a solution $\hat{f}$\ in a
neighborhood of
\begin{equation}
\mathcal{D}_{y}:=I_{\eta,y}^{-}(0)\cap\{y^{0}\geq-\frac{1}{2\lambda}\}\;,
\end{equation} if the initial data are small enough. The Einstein
equations in wave coordinates on $\R_{x}^{n+1}$ have then a solution
on
\begin{equation}
\mathcal{D}_{x}:=\phi^{-1}(\mathcal{D}_{y})\equiv
I_{\eta,x}^{+}(0)\cap x^{0}\geq\lambda+\sqrt{\lambda^{2}+r^{2}}\;,
\end{equation} $\mathcal{D}_{x}$ is the whole future of the
hyperboloid $x^{0}=\lambda +\sqrt{\lambda^{2}+r^{2}}$, which is the
image by $\phi^{-1}$ of the hyperplane $y^{0}=-\frac{1}{2\lambda}$.

The initial data on $M_{y}$ are deduced by the mapping $\phi$ from
the values on the hyperboloid
$$
 H_{x}:=\{x^{0}=\lambda+\sqrt{r^{2}+\lambda^{2}}\}
$$
of the local solution in $\R_{x}^{n+1}$ and its first derivative, if
$H_{x}$ is included in the domain $\mathcal{V}_{x}$ of its
existence.

The hyperboloid $H_{x} $ is the union of two subsets
\begin{equation}
S_{1}:=H_{x}\cap\{x^{0}-r\geq a\}\text{ \ \ and \ \ \
}S_{2}:=H_{x}\cap \{x^{0}-r\leq a\}
 \;.
\end{equation}
The subset $S_{2}$ is included in the Schwarzschild spacetime
region. On the subset $S_{1}$ it holds that
\begin{equation}
\lambda+\sqrt{r^{2}+\lambda^{2}}\geq r+a
\end{equation}
A simple computation shows that $r$ is bounded on $S_{1}$ if
$\lambda>R_{x}$ and $m$ is small, because, using the value of $a$,
one finds that:
\begin{equation}
r\leq\frac{R_{x}(2\lambda-R_{x})+O(m^2)}{2(\lambda-R_{x}-O(m))}\;,
\end{equation} then $x^{0}$ is also bounded on $S_{1}$. This subset
is therefore included in the existence domain $\mathcal{V}_{x}$ of
the local solution with Cauchy data on $M_{x}$, for small enough
Cauchy data.

We deduce from these results that, for small enough Cauchy data
(including small $m)$ on $M_{x}$ the domain $\mathcal{V}_{x}$ of
the solution contains the future of $x^{0}=2\lambda$, up to and
including $H_{x}$.

On $\phi(H_{x}\cap S_{2})$ the initial data $\hat{f}_{2}$ for $\hat{f}$ is
deduced from the Schwarzschild metric in wave coordinates, which is static, we
have
\begin{equation}
\hat{f}_{2}(\overrightarrow{y})=[\Omega^{-\frac{n-1}{2}}(f_{m}\circ\phi
^{-1})](y^{0}=-\frac{1}{2},\overrightarrow{y})\;, \end{equation}
with, using the expression of $f_{m}$%
\begin{equation}
f_{m}\circ\phi^{-1}(\overrightarrow{y})=\frac{\Omega^{n-2}}{|\overrightarrow
{y}|^{n-2}}h(\frac{\Omega}{|\overrightarrow{y}|},\frac{\overrightarrow{y}%
}{|\overrightarrow{y}|})\;.
 \end{equation}
is a smooth function of $y$ (since the origin $\overrightarrow{y}=0$
does not belong to the domain $\phi(H_{x}\cap S_{2})$) if $n\geq3$.
A simple calculation shows that the same is true of
$\frac{\partial\hat{f}}{\partial y^{0}}$ on $\phi(H_{x}\cap S_{2}) $
as soon as $n\geq5$.

On $\phi(H_{x}\cap S_{1})$, the initial data $\hat{f}_{1}$, are
deduced from the values on the uniformly spacelike submanifold
$H_{x}\cap S_{1}$ of the solution in $\mathcal{V}_{x}$\ and its
derivative, by the restriction of $\phi$ to a neighborhood of
$H_{x}\cap S_{1}$, where $\phi$ is a smooth diffeomorphism.

\subsection{Conclusions}

By general methods used in~\cite{FriedrichSchmidt} one can prove
that our global solutions are causally geodesically complete. On
the other hand, completeness of inextendible spatial geodesics can
be established by adapting the arguments given
in~\cite[Prop.~16.2]{LindbladRodnianski}, details can be found
in~\cite{Loizelet:these}.
We have therefore proved the following theorem:

\begin{theorem}
 \label{Tmain1}
Let $n$ be odd and $n\geq5$. Let be given on $\R^{n}$
gravitational data, perturbation of the Minkowski data given by
sets of functions $\overline {f_{\mu\nu}}$ and
$\overline{\frac{\partial}{\partial x^{0}}f_{\mu\nu}}$. Suppose
that these data satisfy the Einstein constraints and the initial
harmonicity conditions and coincide with the Schwarzschild data of
mass $m$ in the wave coordinates of Section~\ref{sSwc} outside a
ball of finite Euclidean radius $R_x$. If these functions are
small enough in $\big(H_{s+1}\times H_{s}\big)\big(B(R_x)\big)$
norm, $s>n/2$,  then the data admit a geodesically complete
Einsteinian development ($\R^{n+1},g)$.
\end{theorem}

The smoothness of $ \hat f$  as a function of $y$ immediately
provides a full asymptotic expansion of $g$ in terms of inverse
powers of $r=|\vec x|$. In fact, we obtain an asymptotic behavior of
the gravitational field as assumed in~\cite{HollandsIshibashi}.

\section{Einstein Maxwell equations}
\label{SEM}

Let us show that the conformal method extends easily to the
electro-vacuum case.

\subsection{The equations}

The Einstein equations with electromagnetic sources \eq{em1} for the
unknowns $f:=g-\eta$ take in wave coordinates the form
\begin{equation}
g^{\alpha\beta}\frac{\partial^{2}f }{\partial x^{\alpha}\partial
x^{\beta}}=F_{E}(g)(\frac{\partial f}{\partial x},\frac{\partial
A}{\partial x})\;,
\end{equation}
where the right-hand-side is a quadratic form in $\partial f$ and
$\partial A$, with coefficients polynomials in $g$ and its
contravariant associate.
  In the Lorenz gauge, $\nabla_{\lambda}A^{\lambda}=0$,
the Maxwell equations take also in wave coordinates the form
\begin{equation}
 \label{Max}
 g^{\alpha\beta}\frac{\partial^{2}A }{\partial
x^{\alpha}\partial x^{\beta}}=F_{M}(g )(\frac{\partial f}{\partial
x},\frac{\partial A}{\partial
x})\;,{}%
\end{equation}
where the right-hand-side is bilinear in $\frac{\partial f}{\partial
x}$ and in $\frac{\partial A}{\partial x}$, with coefficients
polynomials in $g$ and its contravariant associate.
\renewcommand{\val}[1]{\lvert #1\rvert}%
\newcommand{\boxg}[1]{\tilde{\Box}_g#1}%
\newcommand{\der}[0]{\partial}%
\newcommand{\M}[1]{\mathcal{#1}}%
\newcommand{\reff}[1]{$(\ref{#1})$}%
In fact, \eq{Max} reads in detail
\begin{align}
g^{\alpha\beta}\frac{\partial^{2}A_\sigma  }{\partial
x^{\alpha}\partial x^{\beta}}=-(\der_\sigma g^{\mu\alpha})\der_\mu
A_\alpha-g_{\nu\sigma}g^{\mu\alpha}(\der_\mu
g^{\nu\beta})(\der_\alpha A_\beta-\der_\beta A_\alpha)
 \;;
\end{align}
this uses both the harmonic coordinates condition and the Lorenz
gauge condition. One thus has a system of equations for $(f,A)$
for which the previous analysis applies.

According to Corvino~\cite{CorvinoEM},
the existence of a large set of non-vacuum electro-vacuum initial
data which are Schwarzschildian\footnote{It could appear that a
natural generalisation in this context is to consider initial data
which coincide with those for the $(n+1)$--dimensional
Reissner-Nordstr\"om metric outside of a compact set. However, if
the initial surface   topology is $\R^n$, then the global electric
charge is necessarily zero, and one is back in the Schwarzschild
case.} outside a compact set is also valid for the
Einstein-Maxwell constraints; compare~\cite{CutlerWald}.

\subsection{Conclusions}

We have obtained the following theorem:

\begin{theorem}
\label{Tmain2}
 Let $n$ be odd and $n\geq5$. Let be given on $\R^{n}$
Einstein-Maxwell data, perturbation of the Minkowski data given by
sets of functions $\overline {f_{\mu\nu}}$ and
$\overline{\frac{\partial}{\partial x^{0}}f_{\mu\nu}}$,
$\overline{A_{\lambda}}$ and $\overline{\frac{\partial}{\partial x^{0}%
}A_{\lambda}}$, Suppose that these data satisfy the
Einstein-Maxwell constraints and the initial harmonicity and
Lorenz gauge conditions, and coincide with the Schwarzschild data
with mass $m$ in the wave coordinates of Section~\ref{sSwc}
outside a ball of finite Euclidean radius $R_x$. If these
functions are small enough in $\big(H_{s+1}\times
H_{s}\big)\big(B(R_x)\big)$ norm,  $s>\frac{n}{2}$, then the data
admit a  geodesically complete
electro-vacuum Einsteinian development ($\R^{n+1},g,A)$.
\end{theorem}

Similarly to Theorem~\ref{Tmain1}, one obtains a full asymptotic
expansion both of the metric and of the Maxwell field in terms of
inverse powers of $r$ near Scri.

\bigskip

\textsc{Acknowledgements:} The authors wish to thank the Albert
Einstein Institute for hospitality during work on this paper. We are
grateful to H.~Lindblad for bibliographical advice. PTC acknowledges
useful discussions with J.~Corvino.

\bibliographystyle{amsplain}
\bibliography{../references/reffile,%
../references/newbiblio,%
../references/bibl,%
../references/howard,%
../references/myGR,%
../references/newbib,%
../references/Energy,%
../references/netbiblio,%
../references/PDE}
\end{document}